Evaluating the electron density model by applying an imaginary modification

Hui LI[a]*, Meng HE[b,d]* and Ze ZHANG[c]

[a]Beijing University of Technology, Beijing 100124, People's Republic of China

[b]CAS Key Laboratory of Nanosystem and Hierarchical Fabrication, CAS Center for Excellence in Nanoscience, National Center for Nanoscience and Technology, Beijing 100190, People's Republic of China

[c]Zhejiang University, Hangzhou 310014, People's Republic of China

[d]School of Physical Sciences, University of Chinese Academy of Sciences, Beijing 100049, People's Republic of China

Correspondence e-mail: huilicn@yahoo.com, mhe@nanoctr.cn

**Abstract**

A function has been proposed to evaluate the electron density model constructed by inverse Fourier transform using the observed structure amplitudes and trial phase set. The strategy of this function is applying an imaginary electron density modification to the model, and then measuring how well the calculated structure amplitudes of the modified model matches the expected structure amplitudes for the modified correct model. Since the correct model is not available in advance, a method has been developed to estimate the structure amplitudes of the modified correct model. With the estimated structure amplitudes of the modified correct model, the evaluation function can be calculated approximately. Limited tests on simulated diffraction data indicate that this evaluation function may be valid at the data resolution better than 2.5 Å.

# 1. Introduction

In X-ray crystallography, the structure of a crystal is actually the electron density distribution in the crystal. The electron density distribution of a crystal is the inverse Fourier transform of its structure factors. Unfortunately, only structure amplitudes can be derived from the experimentally recorded diffraction intensity data, phases of the structure factors usually cannot be measured directly. Various techniques have been developed to derive the phases and then build the electron density map, such as the direct methods (Karle and Hauptman, 1950; Woolfson, 1987; Woolfson and Fan, 1995), which has been dominating the small-molecule X-ray crystallography. Previously, we have proposed a new strategy to derive the phases for observed structure amplitudes: The correctly retrieved electron density map is identified by an evaluation function (or a figure of merit) from a tremendous number of electron density models, which are constructed by assigning all possible phase sets (within certain accuracy) to the observed structure amplitudes. Two evaluation functions were also proposed for this purpose (Li et al., 2015). Recently, Burla et al. developed a multipurpose figure of merit, MPF, which can also be used in recognizing the quality of a current electron density model (Burla et al., 2017). Here, we put forward another evaluation function (or figure of merit) for identifying the correctly retrieved electron density map. Instead of examining the electron density model itself, this evaluation function assesses the agreement between the calculated and expected structure amplitudes of the modified electron density model.

Indeed, iterative electron density modification methods for structure determination, such as charge flipping (Oszlányi and Sütő, 2004; 2008), have been developed a long time ago, and are gaining their popularity. In the process of structure determination using iterative electron density modification method, the initial electron density model usually is obtained by inverse Fourier synthesis using observed structure amplitudes and random phases. The obtained electron density map is modified in a particular way in each following iterative cycle, and the correct electro-density map may be obtained finally. Oszlányi and Sütő have proposed an R factor to monitor the convergence of charge flipping iterations (Oszlányi and Sütő, 2004). This R factor measures how well the calculated structure amplitudes of the modified current electron density model agree with the observed ones. Inspired by this R factor proposed by Oszlányi and Sütő, we conceived an evaluation function, which hopefully can be used to identify the correct electron density model. The details of this evaluation function will be presented in the following sections.

# 2. Definition and calculation

## 2.1 Notation

EDM: electron density modification.

FT: Fourier transform

IFT: inverse Fourier transform

$\rho_{cryst}$: the electron density distribution in a real crystal.

$\rho_o$: the image of $\rho_{cryst}$ at a certain resolution, which is calculated by IFT using the observed

structure amplitudes and correct phases.

$\rho$: the current electron density model which is calculated by IFT using the observed structure amplitudes and the current trial phase set.

$\rho_o{}^*$: the resultant electron density map after $\rho_o$ has been modified in a particular way.

$\rho^*$: the resultant electron density map after the current electron density model $\rho$ has been modified in a particular way.

$|F_{obs}|$: the observed structure amplitudes.

$F_{cal}$: the calculated structure factors of the current electron density model, $\rho$.

$F_{cal}{}^*$: the calculated structure factors of $\rho^*$.

$F_{o,\,cal}{}^*$: the calculated structure factors of $\rho_o{}^*$.

$\phi$: the current trial phase set for $|F_{obs}|$.

$f_i$: atomic scattering factor of the $i$th atom in the unit cell.

## 2.2 Definition of the evaluation function

Traditionally, the so-called crystallographic residual is generally used to evaluate how well the structure model agrees with the observed structure amplitudes. The crystallographic residual is defined as

$$\text{RES} = \frac{\sum_h ||F_{obs}| - |F_{cal}||}{\sum_h |F_{obs}|}. \quad (1)$$

RES usually can only be used in cases where an approximately correct structural model is available, and cannot tell if a trial phase set $\phi$ is correct or not because $|F_{cal}| = |F_{obs}|$ always holds regardless of whatever the phase set is assigned to the observed structure amplitudes.

An R factor slightly different RES is proposed by Oszlányi and Sütő, and defined as

$$\text{R} = \frac{\sum_h ||F_{obs}| - |F_{cal}^*||}{\sum_h |F_{obs}|}. \quad (2)$$

$|F_{cal}{}^*|$ is the structure amplitudes of the modified current electron density model, therefore is always different from $|F_{obs}|$ even if the model is correct. When the high resolution diffraction data is available and the trial phase set $\phi$ is correct (within certain accuracy), an EDM implemented in a particular way (for example, electron density flipping in low-density regions as implemented in charge flipping algorithm) results in very little difference between $\rho$ and $\rho^*$, therefore a small R factor is an indicator of the correctness of the structure model. Nevertheless, when the resolution of the diffraction data is poor, considerable electron density will locate at intermediate regions between atoms even the electron density model is correctly constructed using the correct phase set. Then an EDM, which usually take effect in low-density regions, will lead to significant difference between $\rho$ and $\rho^*$. We expect that the R factor will no longer be capable of indicating the correct model in such cases. This conjecture was confirmed by examining the evolution of R factor in successive EDM iterative cycles using simulated diffraction data at various resolutions. As shown in Figure 1, with the deterioration of the data resolution, R values for the correct electron density models increase, and so do the finally resultant R factors of iterative EDM cycles. In the case where the data resolution is 2.5 Å, The R value for the correct electron density model is significantly higher than the finally resultant one obtained from the iterative cycles. Clearly, in such a case, R factor defined in equation (2) is no longer capable of indicating the correct electron density model. We also note that R factor works well when the data of high resolution is available and the correct model can be retrieved from the diffraction data by using the iterative EDM method. This implies that it might be unnecessary to find a better figure of merit for iterative EDM algorithms to monitor the evolution of electron density model in the successive iterative cycles, but we do need a new figure of merit different from R factor for identifying the correctly retrieved electron density model when the data resolution is poor.

We propose $R_{Tian}$ factor which is defined as

$$R_{Tian} = \frac{\sum_h ||F^*_{o,cal}|-|F^*_{cal}||}{\sum_h |F^*_{o,cal}|} \quad (3)$$

where $F^*_{cal}$ and $F^*_{o,cal}$ represent the calculated structure factors obtained from the modified current and correct electron density models, respectively, to which the same EDM has been applied. When the trial phase set is correct within certain accuracy, $R_{Tian}$ will approach to zero. This holds true even for low-resolution data. The difficulty lies in that $|F^*_{o,cal}|$ cannot be obtained before the correct electron density model is available. If a reasonable approximation of $|F^*_{o,cal}|$ can be established using the available data, then $R_{Tian}$ can be used as a figure of merit to identify the correct electron density model.

## 2.3 The estimation of $R_{Tian}$ factor

We suppose that the lattice parameters, the constituent atoms and their numbers in the unit cell are known together with the observed structure amplitudes before the structure determination. The number of atoms in the unit cell of the investigated crystal is assumed to be N, and the atomic scattering factor of the $i$th atom is supposed to be $f_i$.

For a hypothetical structure which has the same unit cell as that of the investigated crystal and only one atom with an atomic scattering factor of $f_i$ at the origin, its structure factor can be readily calculated using the following equation

$$F_{i,h} = \sum f_i \times \exp(-2\pi i \vec{h} \cdot \vec{r}) = f_i. \quad (4)$$

The electron density map of the hypothetical structure $\rho_i$ can be calculated by a straightforward IFT, namely,

$$\rho_i = IFT(F_{i,h}) = IFT(f_i). \quad (5)$$

The Fourier synthesis in equation (5) is terminated at a resolution identical to that of the observed data of the investigated crystal. We suppose $\rho_i$ changes into $\rho_i^*$ after applying an imaginary EDM, then the structure factor of $\rho_i^*$, which is denoted as $F_{i,h}^*$, can be obtained by Fourier transform. We suppose the modification applied to $\rho_i$ is $V_i$, then we have

$$\rho_i^* = \rho_i \times V_i \quad (6)$$

and

$$F_{i,h}^* = \text{FT}(\rho_i^*) = \text{FT}(\rho_i \times V_i) = FT(\rho_i) \otimes FT(V_i) = f_i \otimes v_i \qquad (7)$$

where $\otimes$ represents convolution, and $v_i$ represents $FT(V_i)$. In analogy to $\rho_i$, $\rho_i^*$ may be considered as the electron density map built up using a pseudo-atom at the origin with an atomic scattering factor $f_i^* = f_i \otimes v_i$, which is hereinafter referred to as pseudo-atomic scattering factor.

The imaginary EDM applied to $\rho_i$ is assumed to be

$$V_i = \begin{cases} 1, & \rho_i \geq \delta_i \\ 0, & \rho_i < \delta_i \end{cases} \qquad (8)$$

where $\delta_i$ is a real number close to 0. When $\delta_i$ is appropriately selected, $\rho_i^*$ remains the same as $\rho_i$ only in a small region around the center of the atom and is set to zero in the rest regions. An example is given in Figure 2 to illustrate how $\rho_i^*$ and $f_i^*$ vary with $\delta_i$. It can be clearly seen from Figure 2 that positive $\rho_i^*$ is limited to the region very proximate to the atomic position even when $\delta_i$ is quite a small positive value. In addition, it is found that $f_i^*$ varies significantly with $\delta_i$, especially in the low $\sin\theta/\lambda$ range. When $\delta_i$ is set to an appropriate value, $f_i^*$ coincides with the original atomic scattering factor very well in the low $\sin\theta/\lambda$ range, and only in the high $\sin\theta/\lambda$ range significant difference between $f_i^*$ and $f_i$ can be observed. Since $f_i$ value at $\theta = 0$ represents the number of electrons of the $i$th atom, $f_i^*$ coinciding with $f_i$ in the low $\sin\theta/\lambda$ range implies that the pseudo-atom has the same number of electrons as that of the $i$th atom at this $\delta_i$ value.

The "observed" electron density distribution of a crystal, $\rho_o$, may be considered as the superposition of electron density of each constituent atom in the unit cell, which can be represented by

$$\rho_o = \sum_i^N \rho_i(r_i) \qquad (9)$$

where $\rho_i(r_i)$ is $\rho_i$ translated from the origin to the atomic position $r_i$.

We assume

$$\rho_o^* = \rho_o \times \sum_i^N V_i(r_i) = \sum_i^N \rho_i(r_i) \times \sum_j^N V_j(r_j) \qquad (10)$$

where $V_i(r_i)$ is $V_i$ translated from the origin to the atomic position $r_i$.

It is achievable that region with $V_j(r_j) = 1$ of each atom does not overlap with that of adjacent atoms by selecting appropriate $\delta_i$. In such cases,

$$\rho_i(r_i) \times V_j(r_j) = 0 \text{ for } i \neq j.$$

Then we obtain

$$\rho_o^* = \sum_i^N \rho_i(r_i) \times V_i(r_i) = \sum_i^N \rho_i^*(r_i) \qquad (11)$$

and

$$F_{o,cal}^* = FT(\rho_o^*) = \sum_i^N f_i^* \exp(-2\pi i \vec{h} \cdot \vec{r_i}). \qquad (12)$$

The experimental observed intensity of a crystal is given by

$$|F_{obs}|^2 = K \sum_i^N \sum_j^N f_i f_j \exp[2\pi i \vec{h} \cdot (\vec{r_i} - \vec{r_j})]$$

$$= K\{\sum_i^N f_i^2 + \sum_i^N \sum_{j(i \neq j)}^N f_i f_j \exp[2\pi i \vec{h} \cdot (\vec{r_i} - \vec{r_j})]\}$$

$$= K \sum_i^N f_i^2 \left\{ 1 + \frac{\sum_i^N \sum_{j(i \neq j)}^N f_i f_j \exp[2\pi i \vec{h} \cdot (\vec{r_i} - \vec{r_j})]}{\sum_i^N f_i^2} \right\} \quad (13)$$

where K is a scale factor independent of θ. According to equation (13), the intensity of the imaginary crystal with electron density distribution $\rho_o^*$ should be

$$\left| F_{o,cal}^* \right|^2 = K \sum_i^N f_i^{*2} \left\{ 1 + \frac{\sum_i^N \sum_{j(i \neq j)}^N f_i^* f_j^* \exp[2\pi i \vec{h} \cdot (\vec{r_i} - \vec{r_j})]}{\sum_i^N f_i^{*2}} \right\}$$

$$= K \sum_i^N f_i^{*2} \left\{ 1 + \frac{\sum_i^N \sum_{j(i \neq j)}^N k_i k_j f_i f_j \exp[2\pi i \vec{h} \cdot (\vec{r_i} - \vec{r_j})]}{\sum_i^N k_i^2 f_i^2} \right\} \quad (14)$$

where $k_i = f_i^*/f_i$ and $k_j = f_j^*/f_j$. If $k_i = k_j$, then we obtain

$$\left| F_{o,cal}^* \right|^2 = \frac{|F_{obs}|^2}{\sum_i^N f_i^2} \sum_i^N f_i^{*2} \quad (15)$$

As discussed above, $f_i^*$ can coincide with $f_i$ in the low θ range when $\delta_i$ is appropriately set. This implies that in such cases $k_i = k_j = 1$ and equation (15) is valid. At high θ angle, $k_i = k_j$ seems to be a prerequisite hard to meet. Nevertheless, we note that equation (15) agrees well with results derived from Wilson distribution (Wilson, 1942), which is described by

$$\langle |F_{obs}|^2 \rangle = K' \sum_i^N f_i^2 \quad (16)$$

where $\langle |F_{obs}|^2 \rangle$ is the averaged $|F_{obs}|^2$ over a given range of $\sin^2\theta$, and $K'$ is a scale factor independent of θ. Equation (16) is valid when λ/sinθ is small compared with the inter-atomic distances. Similarly, the following equation

$$\langle \left| F_{o,cal}^* \right|^2 \rangle = K' \sum_i^N f_i^{*2} \quad (17)$$

is valid when λ/sinθ is small. Combining equations (16) and (17), we obtain

$$\langle \left| F_{o,cal}^* \right|^2 \rangle = \frac{\langle |F_{obs}|^2 \rangle}{\sum_i^N f_i^2} \sum_i^N f_i^{*2} \quad (18).$$

Although the validity of equation (18) does not mean that equation (15) is valid, it implies that estimating $\left| F_{o,cal}^* \right|^2$ using equation (15) is statistically reasonable.

Combining equations (3) and (15), we obtain

$$R_{Tian} = \frac{\sum_h ||F_{o,cal}^*| - |F_{cal}^*||}{\sum_h |F_{o,cal}^*|} = \frac{\sum_h \left| \sqrt{\frac{|F_{obs}|^2}{\sum_i^N f_i^2} \sum_i^N f_i^{*2}} - |F_{cal}^*| \right|}{\sum_h \sqrt{\frac{|F_{obs}|^2}{\sum_i^N f_i^2} \sum_i^N f_i^{*2}}} \quad (19)$$

Note that EDM applied to both the correct and the current electron density models should be the same in equation (3). In equation (19), we actually have assumed that the multiplier applied to modify the correct electron density model is $\sum_i^N V_i(r_i)$ by taking $\sqrt{\frac{|F_{obs}|^2}{\sum_i^N f_i^2} \sum_i^N f_i^{*2}}$ as the approximation of $|F_{o,cal}^*|$. Therefore, the current electron density model should be modified by multiplying it with $\sum_i^N V_i(r_i)$. Unfortunately, $\sum_i^N V_i(r_i)$ contains the atomic position $r_i$ and cannot be determined prior to the structure determination. An alternative way to modify the current electron density model is multiplying it with

$$W = \begin{cases} 1, & \rho \geq \delta_{cryst} \\ 0, & \rho < \delta_{cryst} \end{cases}$$

where $\delta_{cryst}$ is a real number. Clearly, $|F_{cal}^*|$ varies with both the trial phase set and $\delta_{cryst}$. When the trial phase set is correct and W is close to $\sum_i^N V_i(r_i)$, $R_{Tian}$ calculated with equation (19) reaches the minimum, ideally, zero.

### 3. Discussion

3.1 the validity of the method to estimate $|F_{o,cal}^*|$

The validity of the method to estimate $|F_{o,cal}^*|$ using equation (15) is tested on simulated diffraction data, which is obtained from the known structure of $C_6Br_6$. The crystal structure of $C_6Br_6$ was solved and reported by Wu et al. (Wu et al. 2004). Electron density maps with the resolution of 1.0, 1.5 and 2.5 Å are calculated by IFT. At each data resolution, $\delta_C$ and $\delta_{Br}$ are appropriately set to satisfy the condition that $f_i^*$ coincides with $f_i$ approximately in the low $\sin\theta/\lambda$ region. Then the theoretical values of $|F_{o,cal}^*|$ are calculated using equation (14) and shown in Figure 3 as red crosses. The estimated values of $|F_{o,cal}^*|$ are calculated using equation (15) and presented in Figure 3 as blue squares. As shown in Figure 3, at each data resolution, the estimated $|F_{o,cal}^*|$ values agree well with the theoretical ones. To derive equation (15) from equation (14), we have assumed that atomic scattering factors of different atoms vary by the same proportion as a result of EDM. It seems to be a condition hard to be satisfied. Nevertheless, tests on simulated diffraction data of $C_6Br_6$ indicated that equation (15) is valid in estimating $|F_{o,cal}^*|$ even there are atoms with significantly different scattering power, such as C and Br, in the unit cell. On the other hand, we also would like to explicitly state that equation (15) can only be applied to estimate the structure amplitudes of an electron density map modified in a specific way, which has been described in details in section 2.3.

3.2 the validity of $R_{Tian}$ in identifying the correct phase set

The factor $R_{Tian}$ is aimed to identifying the correct phase set from a great number of possible phase sets assigned to the observed structure amplitudes. The capability of $R_{Tian}$ in evaluating the trial phase sets is tested on simulated diffraction data of known structure $C_{252}H_{326}O_{19}$ reported by Czugler et al. (Czugler et al. 2003). Two data sets at the resolution of 1.0 and 2.5 Å, respectively, are generated. 400 iterative EDM cycles are performed to retrieve the correct electron density model from the simulated data set at the resolution of 1.0 Å. The initial electron density models are constructed by FT using the "observed" structure amplitudes and random phases. In each following iterative cycle, electron densities greater than $\delta_{cryst}$ are retained, and the others are set to zero. The variation of R and $R_{Tian}$ factors with the iterative EDM cycles is shown in Figure 4a. Both R and $R_{Tian}$ drop drastically at the early stage and then keep stable in the following iterative cycles. The convergent R and $R_{Tian}$ values coincide with each other, and both are much higher than the expected values for the correct model. It is noteworthy that the expected value of $R_{Tian}$ is lower than that of R factor. In this case, the correct model cannot be retrieved by iterative EDM cycles, as indicated by the high convergent R and $R_{Tian}$ values. Then a partial structure including 10 correctly located carbon atoms is build up, and a phase set is derived from this partial structure. This phase set is combined with the "observed" structure amplitudes to generate the initial electron density model. Starting with this initial electron density model, iterative EDM cycles converge to the correct electron density map. The evolution of R and $R_{Tian}$ factors in this attempt is shown in Figure 4b. In this case, the convergent R value approaches to the expected value for the correct model, while the convergent $R_{Tian}$ value is still significantly higher than the expected value for the correct model. The discrepancy between the convergent and the expected value of $R_{Tian}$ is due to the difference between the convergent and the correct electron density model. This indicates that $R_{Tian}$ is more sensitive than R to the variation of electron density models. At the resolution of 2.5 Å, we use the correct phase set to establish the initial electron density model for the iterative EDM cycles. The evolution of R and $R_{Tian}$ in this case is shown in Figure 4c. The R factor decreases considerably with the iterative cycles while $R_{Tian}$ increases significantly in the same process. This indicates clearly that R factor is no longer valid at this resolution, but $R_{Tian}$ is still valid for identifying the correct phase set. In addition, we note that $R_{Tian}$ is lower than R for the correct model at both data resolution, and the latter increases faster than the former when the data resolution deteriorates.

Tests are also performed on the simulated data of $C_6Br_6$ at the resolution of 2.5 Å. The variation of $R_{Tian}$ and R with the iterative EDM cycles is shown in Figure 5a. As discussed in section 2.2, the R factor is no longer valid in identifying the correct electron density model in this case. It is interesting to note that $R_{Tian}$ reaches a minimum during the iterative process, which is close to the expected value for the correct electron density model. The electron density model corresponding to the minimum value of $R_{Tian}$ is presented in Figure 5b. Comparison between this model and the correct electron density map (shown in Figure 5c) reveals that it is a close resemblance to the correct map. This example further confirms that $R_{Tian}$ is still valid in identifying the correct electron density model at a resolution of 2.5 Å. Moreover, $R_{Tian}$ seems to be more sensitive than R factor because it varies more significantly than R, especially when an approximately correct model has been obtained.

3.3 Application of $R_{Tian}$ in structure determination

In section 3.2, we have checked the capability of $R_{Tian}$ in identifying correct phase set by investigating the variation of $R_{Tian}$ with the iterative EDM cycles. This does not mean that the $R_{Tian}$ factor is designed for iterative EDM algorithms. Usually, iterative EDM algorithms work only in cases where high resolution diffraction data is available. In such cases, $R_{Tian}$ is only slightly different from R factor numerically, and R factor can be more readily calculated than $R_{Tian}$. $R_{Tian}$ factor is independent of iterative EDM algorithms. For an arbitrary trial phase set assigned to the observed structure amplitudes, $R_{Tian}$ can be estimated approximately. The correct phase set is supposed to have the lowest $R_{Tian}$ value. Tests on simulated diffraction data indicate that $R_{Tian}$ is valid even at the resolution of 2.5 Å. Clearly, since the number of all possible phase sets is tremendous, it is not realistic to estimate $R_{Tian}$ values for all possible phase sets and then pick out the phase set with the lowest $R_{Tian}$ value. Potentially, $R_{Tian}$ may be applied as a goal function of global optimization algorithms for retrieving the electron density map from data set of poor resolution.

**Acknowledgements**

This work was financially supported by the National Natural Science Foundation of China (grant Nos. 11574060 and 11474281). HL and MH are very grateful to Ms. Guang-Lian Tian for her consistent support and encouragement.

**Figures**

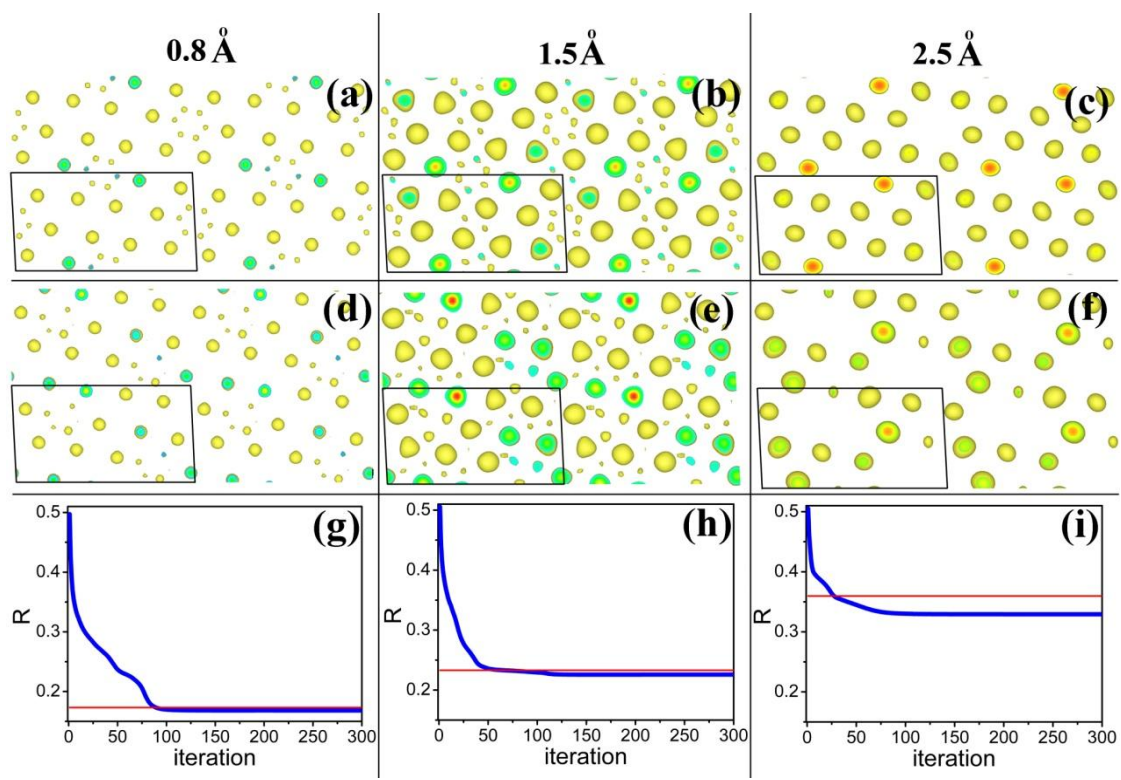

Figure 1. Electron density maps of $C_6Br_6$ at the resolution of 0.8 (a), 1.5 (b) and 2.5 Å (c) constructed by Fourier synthesis using structure amplitudes and correct phases. Shown in (d-f) are the resultant electron density models at the resolution of 0.8, 1.5 and 2.5 Å, respectively, which are obtained from successive iterative EDM cycles. The structure amplitudes and correct phases are calculated from the structure data of $C_6Br_6$ reported by Wu et al. (Wu et al. 2004). All electron density maps in (a-f) are projections of $C_6Br_6$ along [010] direction. The unit cell is represented by a parallelogram in each panel. Note that electron density models shown in (d-f) have origins different from that of the original structural model. Only electron density above a certain positive threshold is illustrated for clarity. (g-i) The evolution of R factor with the iterative EDM cycles. In each iterative cycle, non-negative electron densities are retained and the negative ones are reset to zero. The red horizontal line indicates the R value corresponding to the correct structure model. Panels (g-i) correspond to the iterative cycles using the simulated diffraction data of 0.8, 1.5 and 2.5 Å, respectively. Crystal data of $C_6Br_6$: $P2_1/n$, $a = 8.381$, $b = 4.0192$, $c = 15.3939$ Å, $\beta = 92.674°$, $Z = 2$.

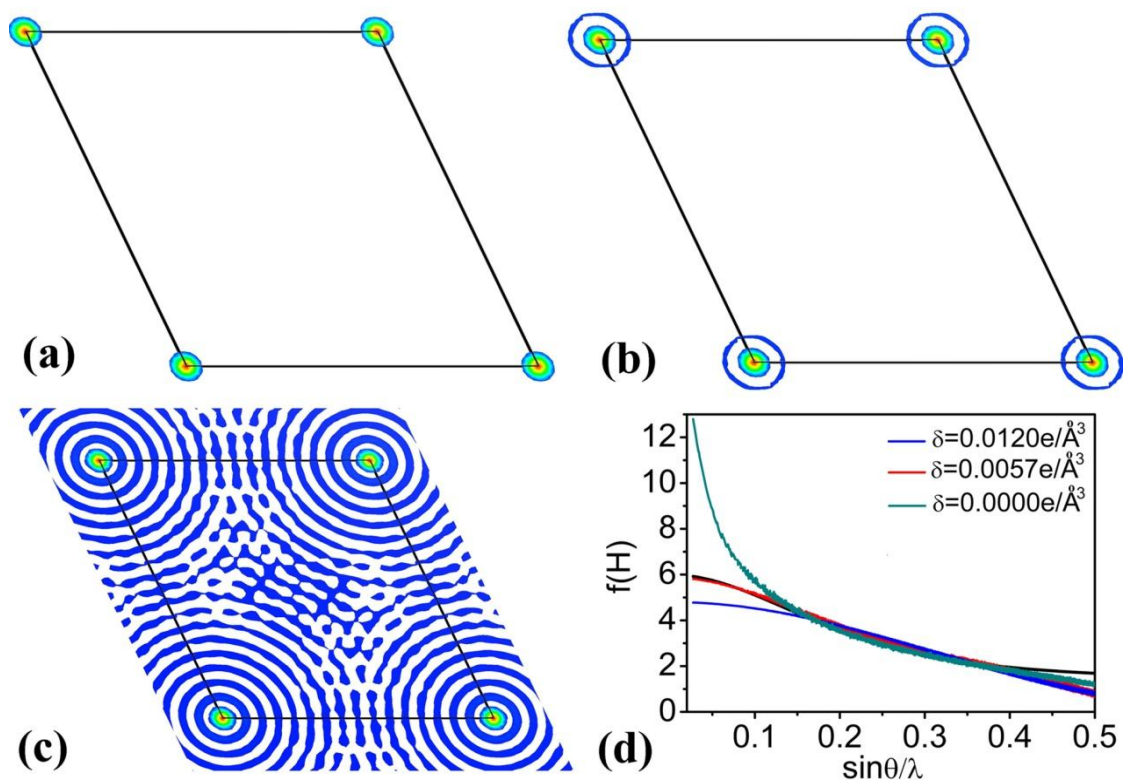

Figure 2. The dependence of $\rho_i^*$ and $f_i^*$ on $\delta_i$. (a-c) the modified electron density distributions on the (001) plane (at $z = 0$) of a hypothetical structure with only one carbon atom at origin. The colored regions have electron densities greater than $\delta_i$. The original electron density map of the hypothetical structure is constructed at the resolution of 1 Å by using equation (5), and then modified according to equations (6) and (8). Maps shown in (a-c) correspond to the resultant electron density distributions modified with $\delta = 0.0$, $0.0057$ and $0.012$ e/Å$^3$, respectively. The hypothetical structure is derived from a known organic compound $C_{252}H_{326}O_{19}$ reported by Czugler et al. (Czugler et al. 2003). (d) $f_i^*$ corresponding to $\delta = 0.0$, $0.0057$ and $0.012$ e/Å$^3$ and the atomic scattering factor of carbon (shown in black).

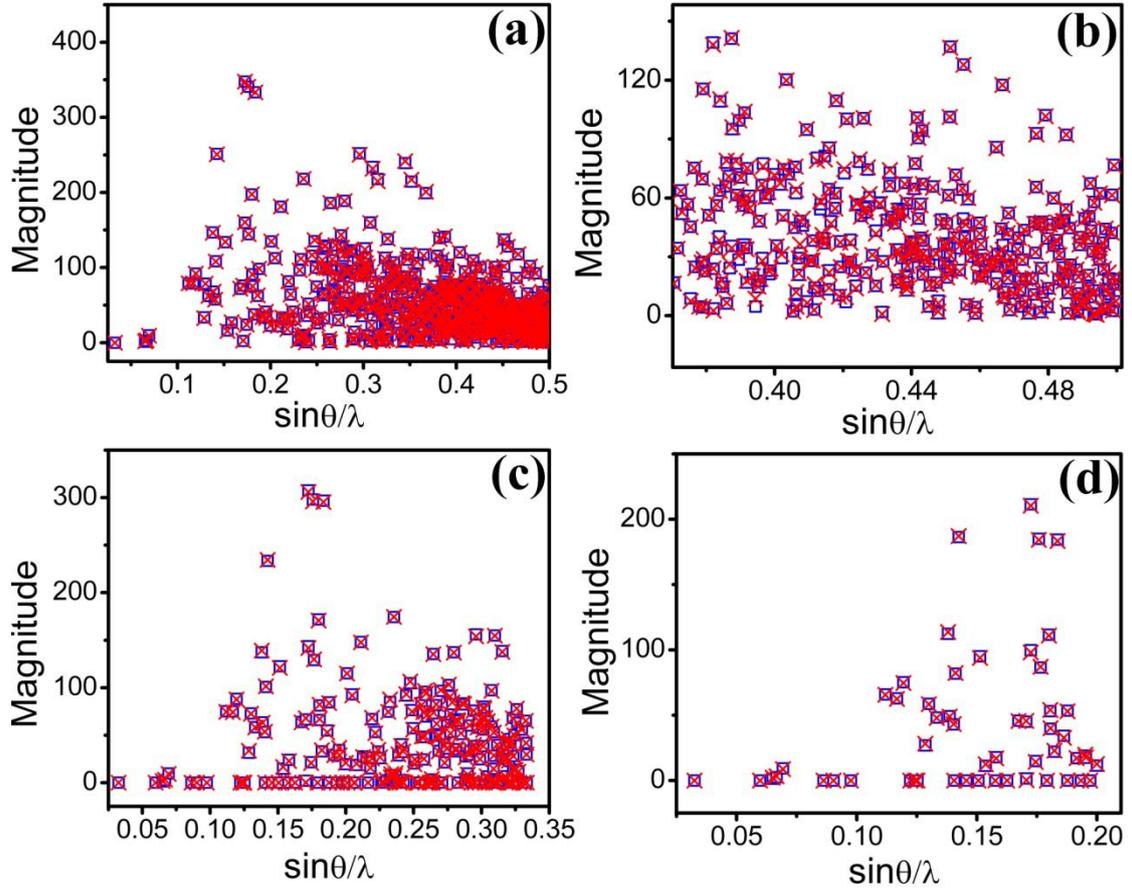

Figure 3. $|F^*_{o,cal}|$ calculated using equations (14) (red cross) and (15) (blue square) of $C_6Br_6$. (a) data resolution: 1.0 Å, $\delta_C = 0.0018$ e/Å$^3$, $\delta_{Br} = 0.125$ e/Å$^3$; (b) is an enlarged view of the bottom-right region of (a). (c) data resolution: 1.5 Å, $\delta_C = 0.003$ e/Å$^3$, $\delta_{Br} = 0.07$ e/Å$^3$; (d) data resolution: 2.5 Å, $\delta_C = 0.0024$ e/Å$^3$, $\delta_{Br} = 0.021$ e/Å$^3$.

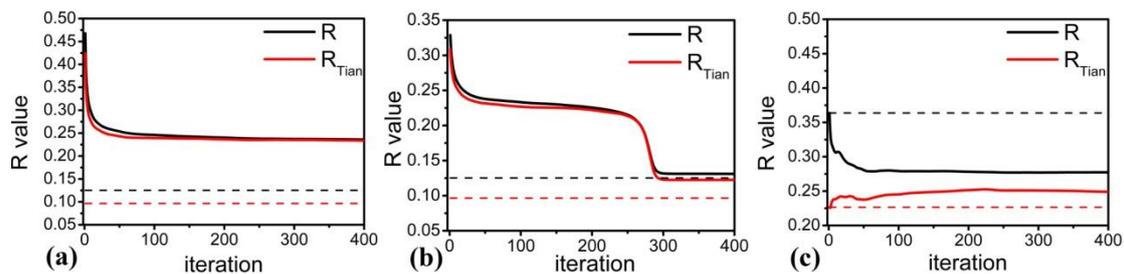

Figure 4. The variation of R and $R_{Tian}$ factors with the iterative EDM cycles at the resolution of 1.0 (a, b) and 2.5 Å (c). The initial electron density model in (a) is generated using random phases while in (b) it is generated using phase set derived from the partial structure. In (c) the correct electron density map of $C_{252}H_{326}O_{19}$ is used as the initial model. In each panel, the target values of R and $R_{Tian}$ corresponding to the correct electron density maps are shown as black and red horizontal dotted lines, respectively. $\delta_{cryst}$ is set to zero in all iterative EDM cycles.

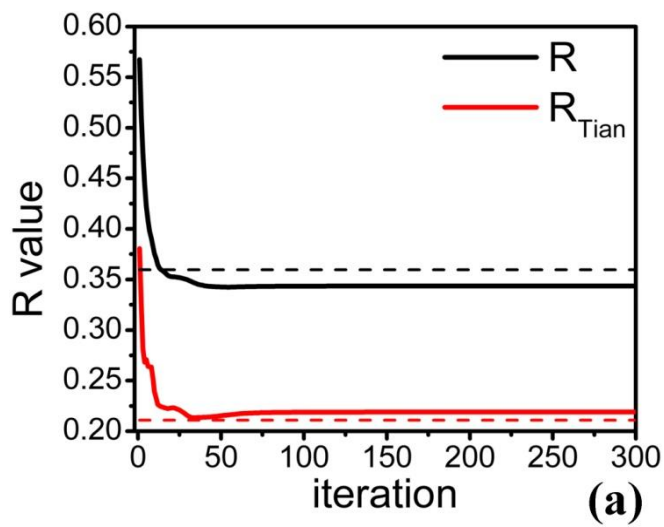

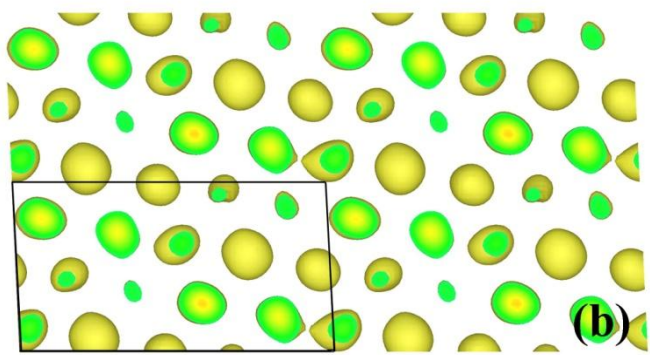

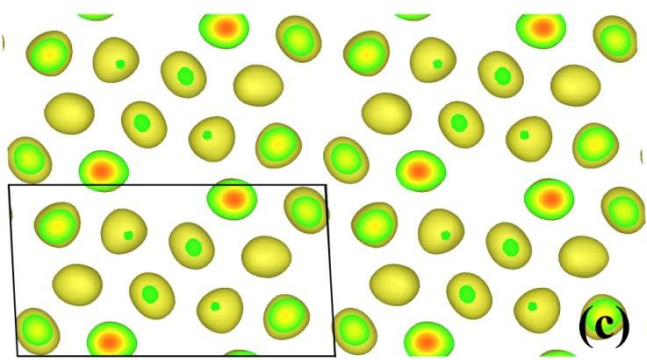

Figure 5. (a) The variation of R and $R_{Tian}$ factor with the iterative EDM cycles at the resolution of 2.5 Å. $\delta_{cryst}$ is set to zero in the iterative EDM cycles. Black and red dotted lines indicate the R and $R_{Tian}$ values for the correct electron density map, respectively. (b) The electron density model corresponding to the minimum value of $R_{Tian}$. (c) The correct electron density map of $C_6Br_6$ at the resolution of 2.5 Å. Both (b) and (c) are projections along [010] direction. Only electron density above 2.5 e/Å$^3$ are illustrated for clarity.